\documentclass[%
 amsmath,amssymb,
 aps,
 twocolumn,
 longbibliography
]{revtex4-2}

\usepackage{graphicx}

\usepackage{bm}

\usepackage[euler]{textgreek}
\usepackage{tikz}
\usepackage{pgf,pgfplots}

\usepackage{hyperref}
\hypersetup{colorlinks=true, allcolors=blue}

\newcommand{\avg}[1]{\left\langle #1 \right\rangle}
\renewcommand{\arraystretch}{1.1}

\pgfplotsset{compat=1.14}

\begin{document}

\preprint{APS/123-QED}
\title{Core polarizability of rubidium using spectroscopy of the $\mathbf{ng}$ to $\mathbf{nh}$, $\mathbf{ni}$ Rydberg transitions}

\author{S.J. Berl}
\author{C.A. Sackett}%
\email{sackett@virginia.edu}
\author{T.F. Gallagher}
\affiliation{%
 Department of Physics, University of Virginia, Charlottesville, Virginia 22904, USA}
 \author{J. Nunkaew}
 \affiliation{Thailand Center of Excellence in Physics, Ministry of Higher Education, Science, Research and Innovation, 328 Si Ayutthaya Road, Bangkok, 10400, Thailand}%

\date{\today}

\begin{abstract}
We present a precise measurement of the rubidium ionic core polarizability. The results can be useful for interpreting experiments such as parity violation or black-body radiation shifts in atomic clocks since the ionic core electrons contribute significantly to the total electrical polarizability of rubidium. We report a dipole polarizability $\alpha_d$ = $9.116 \pm 0.009$~$a_0^3$ and quadrupole polarizability $\alpha_q$ = $38.4 \pm 0.6$ $a_{0}^{5}$ derived from microwave and radio-frequency spectroscopy measurements of Rydberg states with large angular momentum. By using a relatively low principal quantum number ($17 \leq n \leq 19$) and high angular momentum ($4 \leq \ell \leq 6$), systematic effects are reduced compared to previous experiments. We develop an empirical approach to account for non-adiabatic corrections to the polarizability model. The corrections have  less than a 1\% effect on $\alpha_d$ but almost double $\alpha_q$ from its adiabatic value, bringing it into much better agreement with theoretical values. 
\end{abstract}

\maketitle

\section{\label{sec:level1}Introduction}

The electric polarizability of an atom is of significant interest and importance. Accurate polarizability values are needed for many experiments, including atomic clocks, quantum computation, parity-nonconservation, thermometry, and studies of long-range molecules \cite{LeBlanc2007,Mitroy2010,Safronova2011,Dzuba2012,Gaiser2018}. Polarizability measurements are also useful as benchmarks for theoretical calculations since the polarizability depends on the dipole matrix elements of the atomic wave functions, which are difficult to obtain using conventional spectroscopy. Calculation of matrix elements from first principles is  very challenging for multi-electron atoms, so comparisons to experimental quantities, like polarizabilities, provide important checks. These motivations have prompted a series of improving polarizability measurements over the past several decades \cite{Ekstrom1995,Miffre2006,Holmgren2010,Gregoire2015,Leonard2015,Fallon2016,Lee2016,Trubko2017}. One promising new approach is tune-out spectroscopy \cite{Arora2011}, where the ac electric polarizability of an atom vanishes and the wavelength at which that occurs is measured. This technique can provide orders of magnitude improvement in the accuracy of the dipole matrix elements \cite{Leonard2015,Fallon2016,Gregoire2016,Trubko2017}.

Theoretical interpretation of the polarizability is simplest for alkali atoms, where most of the effect comes from the single valence electron. However, the contribution of the core electrons cannot be ignored. For instance, the core contributes about 3\% to the total polarizability of a Rb atom \cite{Safronova2011}, which is large compared to the 0.2\% accuracy of a measurement such as in Ref.~\cite{Gregoire2015}. It can be useful to evaluate and subtract the core contribution from a measurement to obtain the valence polarizability alone, since this provides the most direct connection to the matrix elements of the valence wave functions.  This approach has been used with both dc and tune-out measurements \cite{Leonard2015}, but it is limited by the accuracy to which the core polarizability is known. We present here a new experimental measurement of the core polarizability of Rb, with an accuracy approximately a factor of four better than previously achieved. We expect this to be useful as tune-out spectroscopy and other polarizability measurement techniques continue to improve.

The core polarizability is obtained in our experiment through microwave spectroscopy of atomic Rydberg  states of high orbital angular momentum $\ell$. When the valence electron is far from the core, the atom behaves much like hydrogen; however in any atom other than hydrogen, the Rydberg electron can both penetrate and polarize the ion core, depressing the atomic energy below the analogous hydrogenic energy. Due to the $\ell$ dependent centrifugal barrier keeping the Rydberg electron away from the ion core, Rydberg states of high $\ell$ ($\ell\ge4$ for Rb) have negligible core penetration. However, core polarization remains, and by comparing the energies of the high $\ell$ states to the coressponding hydrogenic energies, the core polarizability can be determined \cite{Born,Edlen}. This method was previously used in Rb with Rydberg states having principal quantum number $n$ in the range of 27 to 30 \cite{Lee2016}.  The accuracy of the spectroscopy measurements was principally limited by Stark shifts from stray dc electric fields. The dc polarizability of a Rydberg atom as a whole is very large, so even fields of 100 mV/cm can be significant \cite{Lee2016}. To address this problem, the work here uses lower principal quantum numbers: $n = 17$ to $19$. Since the atomic polarizability scales as $n^7$, this reduces the electric field sensitivity by a factor of about twenty-five compared to previous work from our group \cite{Lee2016}. Our analysis
combines our new results, previous measurements from \cite{Lee2016},
and recent measurements by Moore {\em et al.} in the range
$n = 38$ to $43$ \cite{Moore20}. 

Because the valence electron produces a non-uniform field at the ion core, the polarization energy of the Rydberg atom depends on both the dipole polarizability $\alpha_d$ and the quadrupole polarizability $\alpha_q$ of the core \cite{Born, Edlen}. Explicitly, the polarization energy is given, in atomic units, 
by \cite{Edlen, Gallagherbook} 
\begin{equation} \label{Wadiab0}
W = -\frac{1}{2}\alpha_d \avg{\frac{1}{r^4}}_{n\ell} - \frac{1}{2} \alpha_q\avg{\frac{1}{r^6}}_{n\ell},
\end{equation}
where $r$ is the distance from the valence electron to the
nucleus, and $ \avg{\frac{1}{r^4}}_{n\ell}$ and $ \avg{\frac{1}{r^6}}_{n\ell}$ are the expectation values of the squares of the expectation values of the field and field gradient at the ion core due to the static probability distribution of the $n\ell$ electron. 
Because of the centrifugal barrier, $ \avg{\frac{1}{r^4}}$ and $ \avg{\frac{1}{r^6}}$ are highly dependent on $\ell$, and measuring the separation between high $\ell$ states of the same $n$ allows us to determine $\alpha_d$ and $\alpha_q$.

Eq.~\eqref{Wadiab0} is often termed the adiabatic core polarization model since it is based on the assumption that the Rydberg electron cloud is static. Of course, the Rydberg electron is not static, and Whitelaw and van Vleck pointed out that 
Eq.~\eqref{Wadiab0} is a limiting case of a shift arising from second order perturbation theory \cite{Whitelaw}. In particular, Eq. (1) is valid when the excited states of the ion lie far above the its ground state compared to the energy spread of the Rydberg states involved. Detailed derivations of Eq.~\eqref{Wadiab0} from the perturbation theory expressions were subsequently given by Mayer and Mayer and van Vleck and Whitelaw \cite{mayer, VanVleck}. Although the adiabatic approximation of Eq.~\eqref{Wadiab0} is reasonably good for Rb, it is not adequate for our purposes. Here we develop an empirical method to account for the nonadiabatic effects. Specifically, we compare Eq.~\eqref{Wadiab0} to the expression of van Vleck and Whitelaw using several simplifications \cite{VanVleck}. This comparison results in correction factors $k_d$ and $k_q$ which must be applied to Eq.~\eqref{Wadiab0}, both of which are unity in the adiabatic approximation. 
Our estimates of
$k_d$ and $k_q$ differ from unity by less than 10\%, and their introduction alters $\alpha_d$ by less than 1\%, but almost doubles $\alpha_q$ from its adiabatic value and brings it into much better agreement with theoretical values \cite{Lee2016,Safronova2011, heinrichs_simple_1970,sternheimer_quadrupole_1970}.

In the sections that follow, we describe the principle and setup of the experiments, the spectroscopic results, the development of the non-adiabatic corrections, the analysis of the polarizablities, and finally our conclusions.

\begin{figure}

\includegraphics{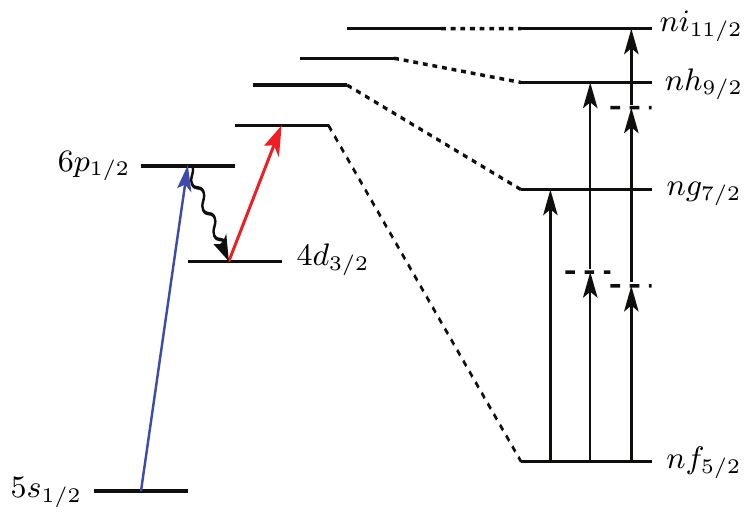}

\caption{\label{fig:states} Atomic states used in this measurement. Rubidium atoms in the $5s_{1/2}$ ground state are optically excited to $6p_{1/2}$, allowed to spontaneously decay to $4d_{3/2}$, and are then optically driven to the $nf_{5/2}$ Rydberg state for $n = 17-19$. The expanded diagram on the right shows microwave transitions from $nf$ to $ng$, $nh$, and $ni$ states using one, two, and three photon excitations, respectively. The $f-g$ interval is about 15~GHz, the $g-h$ interval is about 3~GHz, and the $h-i$ interval is about 1~GHz; the precise values depend on $n$.}
\end{figure}

\section{Experimental Approach}

In order to interpret the energy shifts of the Rydberg state in terms of the core polarizability, it is necessary for the valence electron to remain far from the core at all times. In addition to large $n$, this also requires the use of large angular momentum quantum number $\ell$. Core penetration in a Rydberg state causes its fine structure splitting to differ significantly from that of hydrogen. Such distortions are observed in Rb for $\ell \leq 3$, so we use only states with $\ell \geq 4$. The atoms are excited using the scheme shown in Fig.~\ref{fig:states} where a laser pulse first excites the atoms from the $5s_{1/2}$ ground state to the $6p_{1/2}$ excited state. About a third of the excited atoms spontaneously decay to the long-lived $4d_{3/2}$ state, from which they are excited by a second laser pulse to the $nf_{5/2}$ Rydberg state.  From there, microwave and radio frequency pulses drive transitions to the $ng$, $nh$ and $ni$ states. We use the $g-h$ and $h-i$ intervals to determine the dipole ($\alpha_d$) and quadrupole ($\alpha_q$) polarizabilities of Rb$^+$.

\begin{figure}[t]
\includegraphics{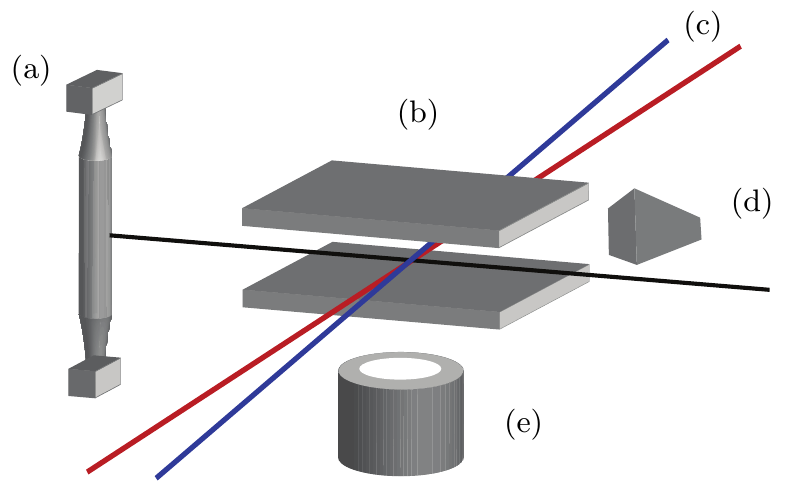}

\caption{\label{fig:setup} Experimental apparatus (not to scale). A rubidium atomic beam is emitted from an oven (a) and passes between electric field plates (b), which are separated by 1.8~cm. Two pulsed laser beams (c) excite the atoms into Rydberg states, and microwaves from the horn (d) drive Rydberg state transitions. An electric field is applied to ionize the Rydberg atoms, and the ions are detected with microchannel plate (e).}
\end{figure}

The experiment is performed in an atomic beam apparatus, shown in Fig.~\ref{fig:setup}. The Rydberg atoms are produced between two electric field plates separated by 1.8~cm. A potential difference of up to 7~kV can be applied between the plates. After the microwave pulse is applied, the electric field is ramped to a value sufficient to ionize the Rydberg states. By carefully controlling the timing and amplitude of the ramp, the atom ionization process can be made state selective such that atoms in $\ell \geq 4$ states are ionized while the $nf$ atoms remain neutral. Any ions produced are detected using a microchannel plate operating in analog mode with spatially integrated channels. The resulting signal current is accumulated using a gated integrator to produce the spectroscopy signal.

The laser excitation pulses are produced by a pair of home-built dye lasers. The first pulse is at a wavelength of 421.5 nm, and is produced using Stilbene 420 dye pumped by the third harmonic of a Quanta-Ray Nd:YAG laser. The second pulse is tuned between 712 nm and 720 nm to populate the desired $nf$ state. This laser uses LD720 dye, pumped by the second harmonic of a Continuum Nd:YAG laser. Both laser pulses have a 20~ns duration, and the second pulse is delayed by 250~ns with respect to the first. Both lasers are linearly polarized perpendicular to the electric field plates. While the $6p$ fine structure is resolved by laser tuning, the $nf$ fine structure is not.

The lifetimes of the $17f$, $18f$, and $19f$ states are $3.2$, $3.8$, and $4.5$~$\mu$s, respectively \cite{Gallagherbook}, and the microwave spectroscopy pulses are applied 1~$\mu$s after the second laser pulse. In the case of the $nf$ to $ng$ transition, a single-photon transition is driven with a microwave frequency ranging from $11$ to $17$ GHz, depending on $n$.  For the $nf$ to $nh$ transition, a two-photon transition is driven with microwaves at half the transition frequency, between 7 and 10 GHz.  For the three-photon $nf$ to $ni$ transition, the two-photon microwave frequency is detuned from the $nh$ state, and we apply a rf frequency near 1~GHz to couple $nh$ to $ni$. These three excitation schemes are illustrated in Fig.~\ref{fig:states}. The microwaves are produced by an Agilent 83622B frequency synthesizer coupled to one of two microwave horns. The rf field is produced by coupling a HP 8673C synthesizer to one of the electric field plates. In all cases, the duration of the spectroscopy pulse is 1~$\mu$s.

\begin{figure}
\includegraphics{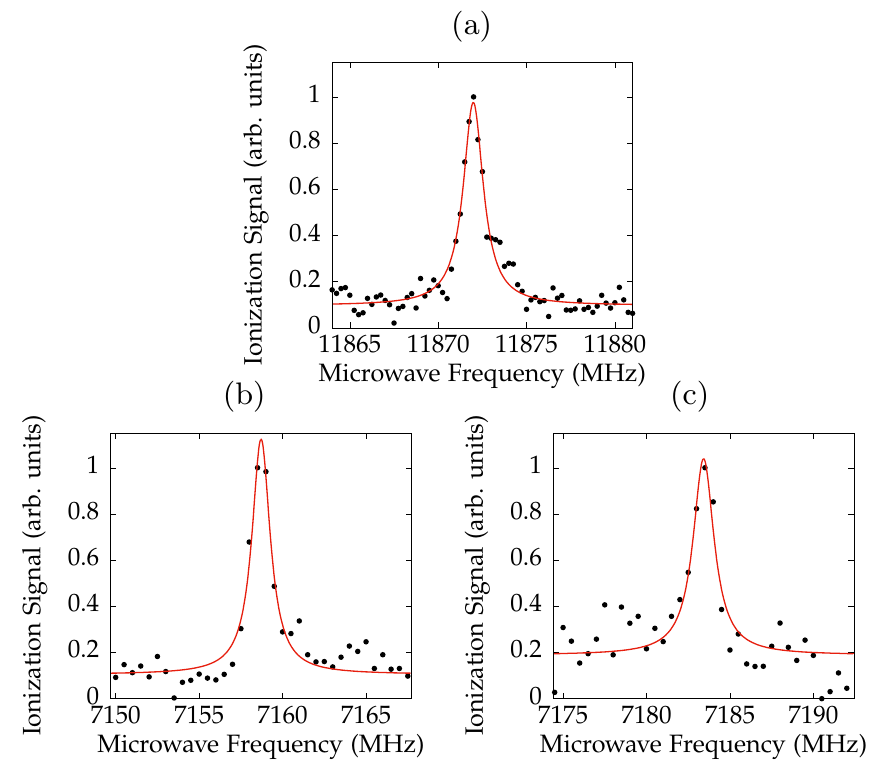}

\caption{ \label{fig:traces} Example spectroscopy line profiles. (a) Single-photon $19f_{5/2} - 19g_{7/2}$.
(b) Two-photon $19f_{5/2}- 19h_{9/2}$. (c) Three-photon $19f_{5/2}- 19i_{11/2}$ with an applied rf frequency of 720~MHz. }
\end{figure}

For each measurement, the microwave frequency is swept across the resonance. The measurement results of each frequency step in the sweep are averaged over ten experimental cycles, and the sweep in its entirety is repeated five times. The resulting signals are averaged to produce a line profile, such as the example data shown in Fig.~\ref{fig:traces}. The profiles are least-squares fit to Lorentzian functions to determine the line centers. Uncertainty in the line center is taken from the uncertainty estimate of the fit. However, in cases where the line center uncertainty from the fit is below 10\% of the fit linewidth, we instead assigned an uncertainty of 10\% of the linewidth to reflect the fact that the actual lineshape is not well characterized.

Several sources of systematic uncertainty must be taken into account, including dc Stark shifts, ac Stark shifts, Zeeman shifts, and fine structure splitting. 

\begin{figure}[t]
\includegraphics{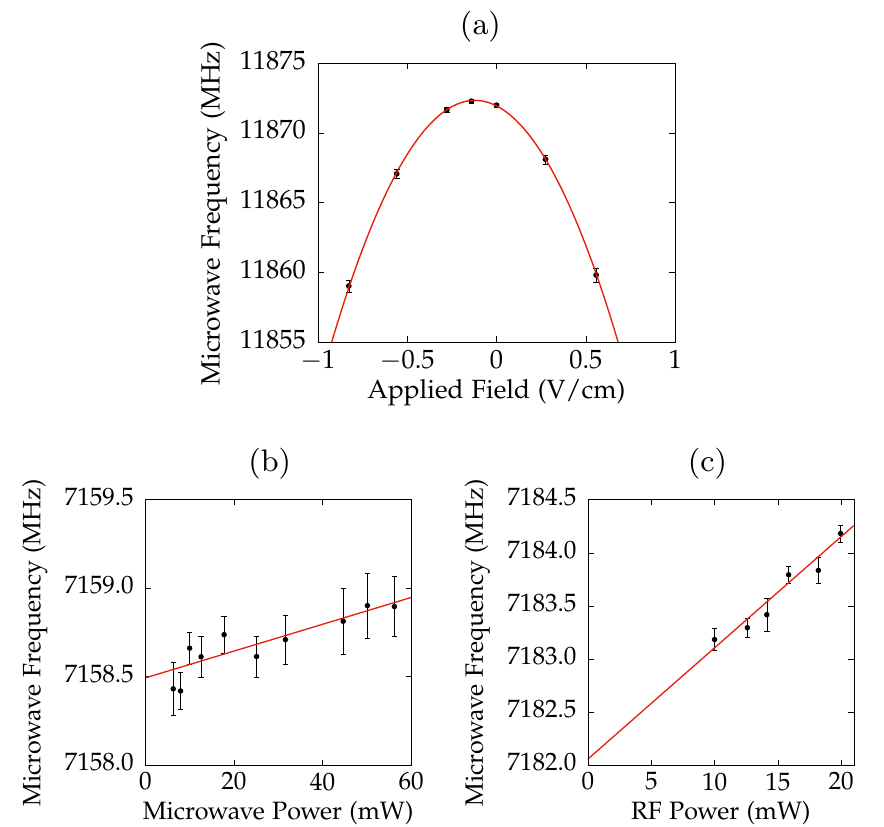}
	
	\caption{\label{fig:stark} 
	a) Measurement of the dc Stark shift on the $19f_{5/2} - 19g_{7/2}$ transition.
	For spectroscopy, the bias voltage is set to the vertex of the curve. 
	b) Measurement of the ac Stark shift
	on the two-photon $19f_{5/2} - 19h_{9/2}$ transition. The line is a linear fit showing the extrapolation to zero microwave power.
	c) Measurement of the ac Stark shift on the three-photon $19f_{5/2} - 19i_{11/2}$ transition. The microwave frequency is swept at a constant 50~mW power. The rf frequency is kept constant at 720~MHz and its power is varied. The line is again a linear fit. } 
\end{figure}

Although dc Stark shifts are reduced by operating at relatively low $n$, they must still be accounted for. The conducting field plates suppress electric fields parallel to the plates, but any residual voltage difference produces a significant field normal to the plates. We are able to apply a bias voltage across the plates during the experiment, and Fig.~\ref{fig:stark}(a) shows how the $nf - ng$ transition frequency varies as a function of the resulting bias field. We fit such data to a parabola and then set the bias voltage to the vertex of the fit. We perform this calibration daily, and observe day-to-day variations of about 0.15~V/cm, corresponding to Stark shifts of the $nf - ng$ transition up to approximately 0.6~MHz.  The apparatus provides no direct way to measure or control the transverse electric field components, but other experiments with similar geometries show that the transverse fields are typically below 0.1 V/cm \cite{Lee2016}. 
We computed the expected Stark shift at this field value and found for $n = 19$ values of 0.25~MHz on the $f-g$ transition, 
0.82 MHz on the $f-h$ transition, and 1.58 MHz on the $f-i$ transition.
These values are in good agreement with the
measured field sensitivity. These shifts are the dominant systematic
uncertainty for the measurement. We scale each shift appropriately with
$n$ and add it in quadrature
to the corresponding experimental frequency error \cite{Zimmerman1979}.  

There are no ac Stark shifts on the single-photon $nf$ to $ng$ transitions, but there are on the multi-photon transitions. These shifts are manifested as linear variations of the transition frequency as a function of microwave or rf power.  We compensate for them by taking data over a range of powers and extrapolating the results to zero power. Example data are shown in Fig.~\ref{fig:stark}(b) and (c). The ac Stark shift is largest for the three-photon $nf - ni$ transition, and the shift depends on the two-photon detuning from the $nh$ state. For these measurements, the microwave and rf powers were independently varied and extrapolated to zero. For each $n$ we used, at least two different two-photon detuning values, with at least one on each side of the $h$ state resonance. The values obtained were consistent with the estimated uncertainties. In all cases, the extrapolation to zero power was performed using an error-weighted least squares fit to the data, and the uncertainty from this fit is reported as the uncertainty in the transition frequency measurement. The resulting values are reported in Table~\ref{tab:transitions}. For the majority of the transitions reported here, at least two measurements were completed on different days, and the results agreed within the stated uncertainty.

The linearly polarized laser beams produce a symmetric distribution of $m$ levels, so we
expect no first-order Zeeman shift. However, the measured background field of about 0.5 G gives a Zeeman energy similar to the fine-structure splitting, so computation of the expected spectrum is complicated.
Instead we applied a bias field comparable to the background field and observed shifts of about 0.1 MHz. We include this as a source of systematic uncertainty in the results.

In hydrogen, the fine structure (FS) splittings of the $n=17-19$ $g$, $h$ and $i$ states range from 1.8~MHz to 0.6~MHz. This is comparable to or less than our experimental linewidth, so the fine structure is not well resolved, but it is significant compared to our measurement accuracy. To avoid uncertainty due to unresolved FS we take advantage of the fact that the excitation scheme of Fig.~\ref{fig:states} ensures that the Rydberg atoms are always in the lower $j$ fine structure state, $j = \ell -1/2$. Accordingly, we have measured the intervals given in Table \ref{tab:transitions}.

\begin{table*}
	\caption{\label{tab:transitions}%
		Measured transition intervals and intervals referenced to the centers of gravity of the fine-structure doublets in MHz for $n=17-19$, $f - g$, $f - h$, and $f - i$. The uncertainties reported correspond to statistical uncertainties added in quadrature with a possible 0.1~V/cm uncontrolled static electric field and an experimentally determined Zeeman shift. 
	}
	\begin{ruledtabular}
		\begin{tabular}{c|cccccc}
			$n$ & $f_{5/2} - g_{7/2}$ & $f_{cg} - g_{cg}$ &  $f_{5/2} - h_{9/2}$ & $f_{cg} - h_{cg}$ & $f_{5/2} - i_{11/2}$ & $f_{cg} - i_{cg}$\\
			\hline
			17 & 16528.7(2) & 16547.3(2) & 19929.5(5) &  19947.8(5) & 20992.5(10) &  21010.6(10)\\
			18 & 13945.2(2) & 13960.9(2) & 16815.6(6) & 16831.0(6) &  17713.2(12) &  17728.5(12) \\
			19 & 11872.3(3) & 11885.7(3) &  14317.0(8) & 14330.2(8) &  15082.9(18) &  15096.0(18) 
		\end{tabular}
	\end{ruledtabular}
\end{table*}

We expect the $\ell\ge4$ FS splittings in Rb to be similar to those of hydrogen, because the $\ell \geq 4$ states should not penetrate the core and the core polarization effect is independent of $j$. To verify this, we retuned the initial laser excitation pulse to the $6p_{3/2}$ state, which then allowed excitation of both the $nf_{5/2}$ and $nf_{7/2}$ states. The $nf_{7/2} - ng_{7/2}$ transition is suppressed due to small Clebsch-Gordan coefficients, but we observed the $nf_{7/2} - ng_{9/2}$ transition.  Using the known $f$-state FS splitting \cite{Han2006}, we obtained a value for the $17g$ FS splitting of $1.83\pm 0.06$ MHz. This is in agreement with the hydrogenic value of $1.78$~MHz and supports the conclusion that the $g$-states are non-penetrating \cite {Tran}. We therefore use the hydrogen FS values for the $\ell \geq 4$ states. For the analysis described below, we use transition frequencies from which the FS shift has been removed by referencing the transition to the center of gravity of the FS manifold. These frequencies are also listed in Table \ref{tab:transitions}.

\section{Analysis \& Discussion}

In our analysis we assume that core penetration does not occur in  Rb states of $\ell\ge 4$ and that relativistic and exchange effects are negligible. In the adiabatic core polarization model the electric field from the quasistatic charge distribution of the Rydberg $n\ell$ electron polarizes the ion core, which results in the polarization energy shift of the Rydberg $n\ell$ state relative to the hydrogenic $n\ell$ energy. The shift is given
by Eq.~\eqref{Wadiab0}, which we here rewrite as 

\begin{equation} \label{Wadiab}
W = -\frac{1}{2}\alpha_d^{(a)} \avg{\frac{1}{r^4}}_{n\ell} - \frac{1}{2} \alpha_q^{(a)}\avg{\frac{1}{r^6}}_{n\ell},
\end{equation}
where the superscripts denote the use of the adiabatic approximation.
If we assume the $n\ell$ wavefunctions to be hydrogenic, there are closed form expressions for the
required expectation values \cite{Edlen,Gallagherbook,Bethe}. As a result, it is a straightforward matter to extract $\alpha_d^{(a)}$ and $\alpha_q^{(a)}$ from the high-$\ell$ Rydberg energies.

Equation~\eqref{Wadiab} gives the energy shift of a state relative to the corresponding state of hydrogen, but we do not have accurate values for the absolute energies of the $nf$ states, so we cannot evaluate the energies of the high-$\ell$ states relative to hydrogen. Instead, we consider the energy difference between two states $n\ell$ and $n\ell'$. Since the hydrogenic energies are independent of $\ell$, the energy difference is
\begin{equation} \label{DWadb}
\Delta W = -\frac{1}{2}\alpha_d^{(a)} \Delta_d^{(a)}
- \frac{1}{2}\alpha_q^{(a)} \Delta_q^{(a)},
\end{equation}
where
\begin{equation} \label{deltada}
\Delta_d^{(a)} \equiv  \avg{\frac{1}{r^{4}}}_{n\ell}
- \avg{\frac{1}{r^{4}}}_{n\ell'}
\end{equation}
and
\begin{equation} \label{deltaqa}
\Delta_q^{(a)} \equiv  \avg{\frac{1}{r^{6}}}_{n\ell}
- \avg{\frac{1}{r^{6}}}_{n\ell'}
\end{equation} 
 The energy difference $\Delta W$ corresponds to the FS-corrected transition frequencies reported in Table \ref{tab:transitions}.

\begin{figure}
     \hspace*{-1em}
\includegraphics{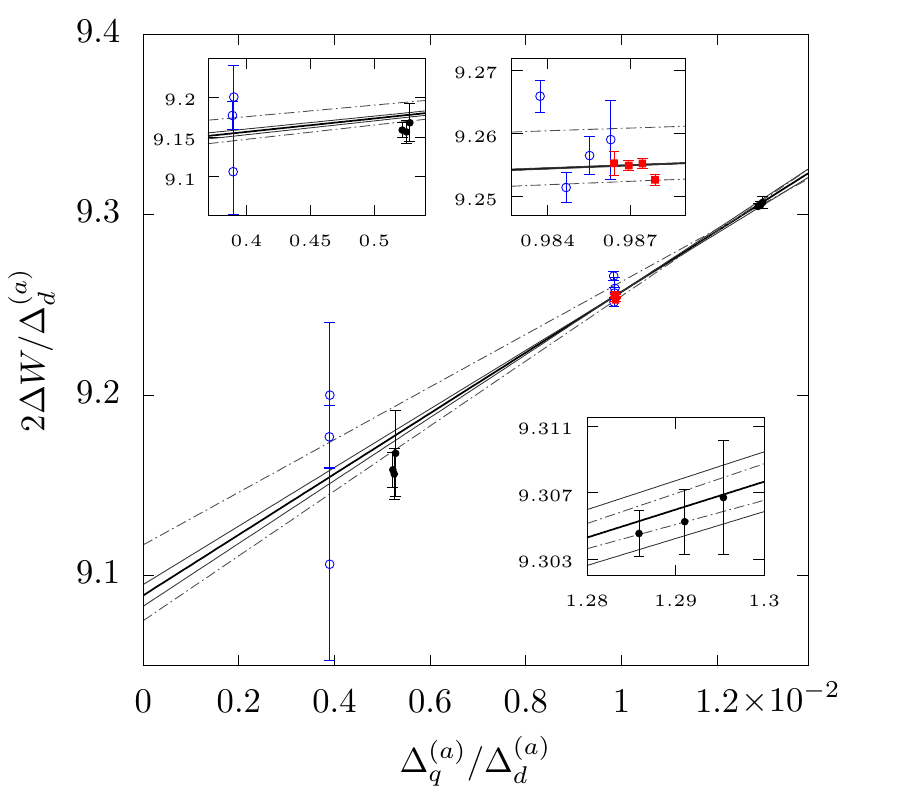}
	\caption{(color online) Core polarizability analysis in the adiabatic approximation. 
	Solid black circles correspond to measurements reported here, open blue circles are values
	from Lee~{\em et al.}~\protect\cite{Lee2016}, and solid red squares are
	from Moore~{\em et al.} The axis quantities are discussed
	in the text, and given in atomic units ($a_0^3$ for the vertical axis, and $a_0^{-2}$ for the horizontal axis). The heavy solid line is a linear fit to all the data shown, weighted by the individual error bars. The thin solid lines show the effect of varying
	the slope and intercept across their $1\sigma$-confidence interval. The dashed
	lines show the confidence interval obtained using only the Berl and Lee data. 
	The intercept and slope of the line give,
	respectively, the dipole and quadrupole polarizabilities in the adiabatic approximation. 
    Insets show expanded views near each set of points.}
    \label{fig:adiabatic}
\end{figure}

Figure~\ref{fig:adiabatic} is a plot of $2\Delta W/\Delta_d^{(a)}$ vs. $\Delta_q^{(a)}/\Delta_d^{(a)}$. The solid black circles show our values for 
$(\ell,\ell')$ pairs $(4,5)$ and $(5,6)$. We also include results from Lee {\em et al.}
\cite{Lee2016} as open blue circles, and from Moore {\em et al.} \cite{Moore20} as solid red squares. 
Lee measured the energy shifts $W$ of individual states relative to hydrogen, so 
for those points the $x$ coordinates are
$\langle r^{-6} \rangle_{n\ell}/\langle r^{-4} \rangle_{n\ell}$
and the $y$ coordinates are $2W/\langle r^{-4} \rangle_{n\ell}$. 
Moore measured the transition frequencies between different
$n$ levels, so the hydrogenic contributions $-R/n^2$ are subtracted
in $\Delta W$, where $R$ is the mass-adjusted Rydberg
constant for $^{85}$Rb.

To obtain estimates for the adiabatic polarizabilities, 
we fit the data points to a line, with the results shown in Table \ref{tab:adiabatic}.
The heavy line in the figure shows the best fits to all the data.
The other lines illustrate the uncertainty range of the fits.
The fit uncertainties 
are calculated differently than by Lee {\em et al.}: here
we estimate the error in a parameter as the change required to increase
$\chi^2/$dof by one when $\chi^2/\text{dof} < 1$, or to double $\chi^2$ when 
$\chi^2/\text{dof} \geq 1$. In Ref.~\cite{Lee2016}, the errors were determined by
increasing $\chi^2/\text{dof}$ by one in all cases. For the Lee data alone,
we now obtain $\alpha^{(a)}_d = 9.12(4)$ and $\alpha^{(a)}_q = 14(4)$, with
$\chi^2/\text{dof} = 4$. In comparison, Lee~{\em et al.}\ 
reported $\alpha^{(a)}_d = 9.12(2)$ and $\alpha^{(a)}_q = 14(3)$.
Note that the low value for $\chi^2$/dof
for the present data alone reflects the fact that our uncertainty is dominated by systematic error
from horizontal bias fields.

It is apparent from the table and the graph that the different data sets are not entirely consistent.
In particular, the $h-i$ frequencies measured here yield points that lie below the overall
best-fit line. The other points involve $g$-state measurements, except for the Lee results at 
low $\Delta_q^{(a)}/\Delta_d^{(a)}$ which are relatively imprecise. A possible explanation
is that the stray fields in our experiment are larger than estimated, which would tend to
decrease the measured $h-i$ intervals as seen. Alternatively, the core-penetration
effect in the $g$ states may be larger than expected. This would tend to increase the measured 
values for those states, which is also consistent. 
Our error analysis accounts for this tension by increasing
the fit uncertainty in the parameters as explained above. 

\begin{table}
	\caption{\label{tab:adiabatic}%
		Calculated values of the dipole and quadrupole polarizabilities, using the adiabatic
		core polarization model. Data sets are Berl (present work), Lee \protect\cite{Lee2016},
		and Moore \cite{Moore20}. Values in parentheses are the estimated uncertainties. The goodness-of-fit parameter $\chi^2$/dof is calculated as
		the sum of the squares of the deviations between the measured data and the fit, divided 
		by the total number of data points used, minus two.
	}
	\begin{ruledtabular}
		\begin{tabular}{l|ccc}
			Data sets & $\alpha_d^{(a)}$ & $\alpha_q^{(a)}$ &  $\chi^2$/dof \\ \hline
			Berl & 9.059(18) & 19.1(1.4) & 0.05 \\
			Berl, Lee & 9.096(21) & 16.3(1.7) & 2.9 \\
			Berl, Lee, Moore & 9.089(6) & 16.8(6) & 2.7 
		\end{tabular}
	\end{ruledtabular}
\end{table}

However, the adiabatic approximation is inadequate here and must be corrected to incorporate non-adiabatic effects \cite{mayer, VanVleck,Freeman,Opik,Merkt}. The non-adiabatic correction arises because Eq.~\eqref{Wadiab} is an approximation to the second-order shift from the multipole expansion of the Coulomb interaction between the Rb$^+$ ion core and the Rydberg electron. The same method of analyzing the experimental data can be used if we introduce correction factors $k_{d,n\ell}$ and $k_{q, n\ell}$ into Eq.~(1), which then reads \cite{Gallagherbook,GallagherBa}
\begin{equation} \label{Wnonad}
    W = -\frac{1}{2} k_{d,n\ell} \alpha_d \avg{\frac{1}{r^4}}_{n\ell}
    -\frac{1}{2} k_{q,n\ell} \alpha_q \avg{\frac{1}{r^6}}_{n\ell}.
\end{equation}

To develop an estimation for $k_d$ and $k_q$. we consider the contribution of the dipole polarizability to the polarization shift of a Rb $n\ell$ state \cite{VanVleck, GallagherBa}. The atomic wavefunction is taken to be a direct product of the ion wavefunction and a hydrogenic wavefunction for the Rydberg electron. Consequently, the total energy is simply the sum of the ion and Rydberg energies. In a bound Rb $n\ell$ state, the Rydberg electron is coupled to the ground $4p^6$ state of Rb$^+$, which we denote as $a$, so the bound Rydberg state is denoted $an\ell$. Similarly, a Rydberg $n'\ell'$ electron coupled to an excited state $b$ of Rb$^+$ is denoted $bn'\ell'$. We restrict our attention to ion states which are dipole coupled to the ground state. In the Rydberg atom, the $a n\ell$ state is coupled by the dipole term of the Coulomb expansion to the $bn'(\ell-1)$ and $bn'(\ell+1)$ states, as well as the $b\epsilon' (\ell-1)$ and $b\epsilon' (\ell+1)$ continua. The resulting second-order dipole energy shift of the $an\ell$ state is given explicitly by \cite{VanVleck,Gallagherbook}
\begin{align} \label{Wng_non}
\Delta W_{d,n\ell} = \frac{1}{3}&\sum_{b,n'} \Bigg[ 
\frac{\ell\langle a|r_1|b\rangle^2 \langle n\ell | r_2^{-2} | n'(\ell-1)\rangle^2}{(2\ell+1)(W_{an\ell}-W_{bn'(\ell-1)})} \nonumber \\
& + \frac{(\ell+1)\langle a|r_1|b\rangle^2 \langle n\ell | r_2^{-2} | n'(\ell+1)\rangle^2}{(2\ell+1)(W_{an\ell}-W_{bn'(\ell+1)})}
\Bigg],
\end{align}
where the sums are understood to include the continua above the Rydberg and ion limits. Here $r_1$ represents a core electron and $r_2$ the Rydberg electron. The $r_2^{-2}$ matrix elements are computed using Numerov's method, and their accuracy is verified using the sum rule \cite{VanVleck}
\begin{equation} \label{complete}
\langle n\ell|r^{2s}|n\ell\rangle =\sum_{n'}\langle n\ell|r^s|n'\ell'\rangle^2.
\end{equation}

\begin{figure}
    \includegraphics[width=\columnwidth]{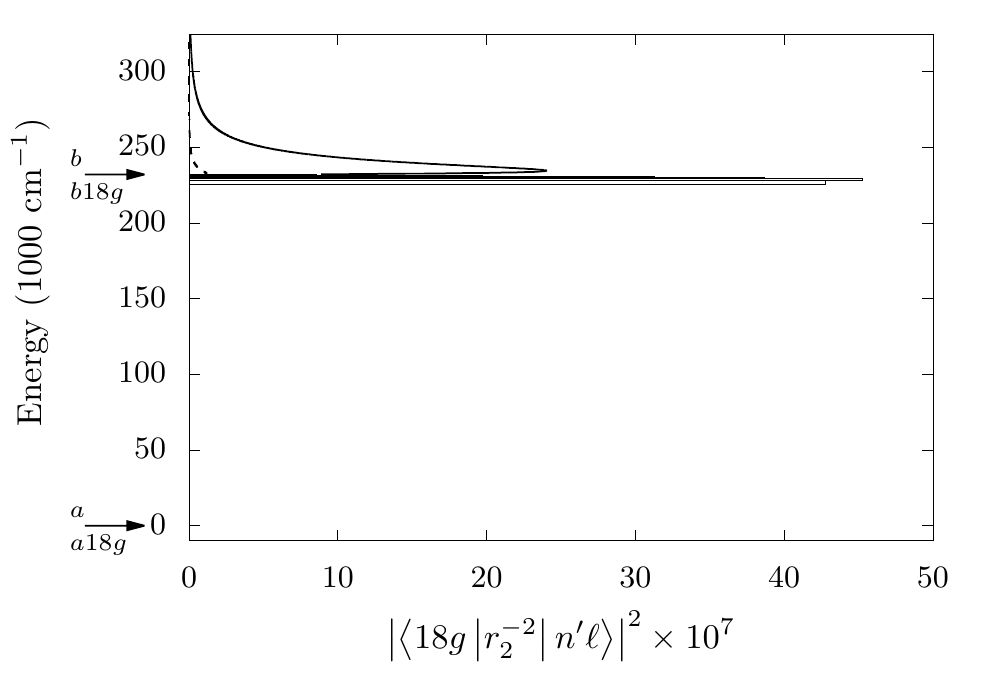}
    \caption{Energy distribution of the $18g$ Rydberg matrix elements. The vertical axis is the energy
    of the Rydberg atom, with the core ion ground state $a$ and excited state $b$ indicated by arrows.
    The energy difference between the ion states and the Rydberg atom states $a18g$ and $b18g$ is too small to resolve. The horizontal axis shows the square matrix elements to the $bn'f$ and
    $bn'h$ states, and also to the $b\epsilon' f$ and $b\epsilon' h$ continua. The continuum states are normalized per unit energy and the bound states are plotted as boxes normalized per unit energy.
    Values for the $h$ states are shown with dashed lines, but the $f$ state matrix elements are generally much larger. The energy range over which the matrix elements remain appreciable is seen to be small, but not very small, compared to $W_b-W_a$. Here we take $W_b = 232\ 300$~cm$^{-1}$,
    corresponding to the effective ion excitation energy $W_{Id}$ discussed in the text.
}
    \label{fig:elements}
\end{figure}

The energy denominators of Eq.~\eqref{Wng_non} can be rewritten as 
\begin{equation} \label{denom}
W_{an\ell}-W_{bn'\ell'}=W_a-W_b+W_{n\ell}-W_{n'\ell'}
\end{equation}
The adiabatic expression of Eq.~\eqref{Wadiab} is the result of taking $W_{n\ell}-W_{n'\ell'}=0$, 
since it is much smaller than $W_b-W_a$.
However, the squared $\langle n\ell|r_2^{-2}|n'\ell'\rangle$ matrix elements
cover a substantial energy range, as shown by the $18g$ example in Fig.~\ref{fig:elements}.
Here the matrix elements cover an energy range that is about 15\% of $W_b-W_a$. The energy range does not depend strongly on 
the Rydberg state energy, so we expect the $k_d$ coefficients in 
Eq.~\eqref{Wnonad} to largely independent of $n$. 

Rather than neglecting $W_{n\ell}-W_{n'\ell'}$ entirely, we consider Taylor expanding
Eq.~\eqref{Wng_non} with  $|W_{n\ell}-W_{n'\ell'}|/|W_a-W_b|$ as a small parameter. To first order, it is possible to show that the sum over the ionic core transitions can be replaced by an effective transition to a single ion state at energy $W_{Id}$ above the ground state, with $W_{Id}$ given by
\begin{equation} \label{Wgi}
\frac{1}{W_{Id}} = \frac{\displaystyle \sum_b \frac{\langle a| r_1|b \rangle^2}{(W_a-W_b)^2}}{\displaystyle \sum_b\frac{\langle a |r_1|b \rangle^2}{W_a-W_b}},
\end{equation}
which is an appropriately weighted average of $1/(W_a-W_b)$. 
Similarly, we can obtain an effective matrix element
\begin{equation}
\langle a|r_1|I \rangle^2 
= \frac{\displaystyle\left(\sum_b \frac{\langle a| r_1|b \rangle^2}{W_a-W_b}\right)^2}%
    {\displaystyle \sum_b \frac{\langle a| r_1|b \rangle^2}{(W_a-W_b)^2}}.
\end{equation}

Replacing the sum over the excited states of the ion with the effective state $I$ allows the ion dipole matrix element to be removed from the sum, leaving
\begin{align} \label{Wng_non2}
\Delta W_{d,n\ell} = \frac{1}{3}\langle a|r_1|I \rangle^2 & \sum_{n'} \Bigg[ 
\frac{\ell \langle n\ell | r_2^{-2} | n'(\ell-1)\rangle^2}{(2\ell+1)(W_{an\ell}-W_{In'(\ell-1)})} \nonumber \\
& + \frac{(\ell+1) \langle n\ell | r_2^{-2} | n'(\ell+1)\rangle^2}{(2\ell+1)(W_{an\ell}-W_{In'(\ell+1)})}
\Bigg],
\end{align}
In practice, it is not necessary to evaluate $\langle a|r_1|I \rangle^2$ since
in this approximation, the ion polarizability is itself simply
$\langle a|r_1|I \rangle^2/6W_{Id}$. 

We do need to determine $W_{Id}$, which requires a knowledge of the distribution of oscillator strength $f_a$ from the ion ground state. Unfortunately, this is not well known. However, the photoionization cross section, proportional to $df_a/dW$, is known and similar to the photoionization cross section of the isoelectronic neutral Kr \cite{Babb,Samson}. For Kr the oscillator strengths are known for both the bound states and the continuum \cite{Berkowitz}, and using them we computed $W_{Id}$ for Kr. We find a value 6\%  higher in energy than the first ionization limit of Kr at 112~900~cm$^{-1}$. We estimate the value for Rb$^+$ to also be 6\% higher than the ionization limit at 220~100~cm$^{-1}$, resulting in  $W_{Id}=232~300$~cm$^{-1}$.

Using $\langle a|r_1|I\rangle^2/3=2\alpha_dW_{Id}$, we can obtain an expression for $k_d$ as 
\begin{align} \label{kd}
k_{d,n\ell} = & \frac{W_{Id}}{\langle n\ell|r_2^{-4}|n\ell\rangle} \sum_{n'} \Bigg[ 
\frac{\ell \langle n\ell | r_2^{-2} | n'(\ell-1)\rangle^2}{(2\ell+1)(W_{an\ell}-W_{In'(\ell-1)})} \nonumber \\
& \qquad + \frac{(\ell+1)\langle n\ell | r_2^{-2} | n'(\ell+1)\rangle^2}{(2\ell+1)(W_{an\ell}-W_{In'(\ell+1)})}
\Bigg].
\end{align}
The values of $k_d$ computed in this way are given in Table \ref{tab:kfacts}.

To obtain an estimate of the uncertainty in $k_d$, we note that $W_{Id}$ is roughly bounded by the
lowest ionic excited state energy and the second ionization energy. For instance, a calculation
of $W_{Id}$ in atomic hydrogen gives a value just above the $1s-2p$ transition energy, which reflects the fact that this transition contains over half of the total oscillator strength. 
In contrast, neutral Kr has six times as much oscillator strength in the first 20 eV above the ionization 
limit as in the bound states \cite{Berkowitz}, which explains 
why $W_{Id}$ is comparable to the ionization energy in that case. The first excited state of Rb$^+$ lies at 
134~000~cm$^{-1}$, about 40\% below the ionization limit. This sets the scale for the uncertainty range,
but we believe the isoelectronic analogy to Kr to be reasonably sound, so we estimate an uncertainty
of $\pm 10\%$ for $W_{Id}$. This translates directly to a 10\% uncertainty in $(1-k_{d, n\ell})$ and provides the uncertainties shown in Table \ref{tab:kfacts}.

The quadrupole correction factor  $k_{q,n\ell}$ is calculated in much the same way as $k_{d,n\ell}$. In this case the $\langle n\ell |r_2^{-3}| n'\ell' \rangle$ matrix elements are required, and they are similarly evaluated numerically for hydrogenic wave functions. To assign an effective energy $W_{Iq}$ accounting for the ionic quadrupole transitions, we use an expression analogous to Eq.~\eqref{Wgi}. Lacking better information, we calculate $W_{Iq}$ for hydrogen
and obtain $122~465$ cm$^{-1}$, which is 12\%  over the ionization limit. The ground state of Rb$^{++}$ is split by the spin-orbit interaction, so we use the center of gravity of the spin-orbit split state as the Rb$^+$ limit. Assuming $W_{Iq}$ to lie 12\% above this results in $W_{Iq}=248~000$~cm$^{-1}$.  Using this value of $W_{Iq}$ in the quadrupole analog of Eq.~\eqref{kd}, we calculate $k_{q,n\ell}$. 
Since there is no analog to the Kr oscillator strength distribution for comparison, we 
assign a $\pm 20\%$ uncertainty to $W_{Iq}$ and thus to $1-k_{q,n\ell}$.
The results are also shown in Table \ref{tab:kfacts}.

\begin{table}
    \centering
        \caption{Non-adiabatic correction factors, calculated as in Eq.~\protect\eqref{kd}. The lower-$n$ values are relevant to the data taken here, and the higher-$n$ values are for the data of Ref.~\protect\cite{Lee2016}.}
    \label{tab:kfacts}
    \begin{ruledtabular}
    \begin{tabular}{cccc|ccc}
    & & $k_d$ & & & $k_q$ &  \\
$n$ & $\ell = 4$ & $5$ & $6$ & $4$ & $5$ & $6$ \\ \hline
17-19 & 0.978(2) & 0.990(1) & 0.994(1) & 0.919(15) & 0.966(7) & 0.984(3) \\
27-42 & 0.977(2) & 0.990(1) & & 0.919(15) & 0.966(7) & \\
\end{tabular}
\end{ruledtabular}
\end{table}

\begin{figure}
    \centering
    \includegraphics[width=\columnwidth]{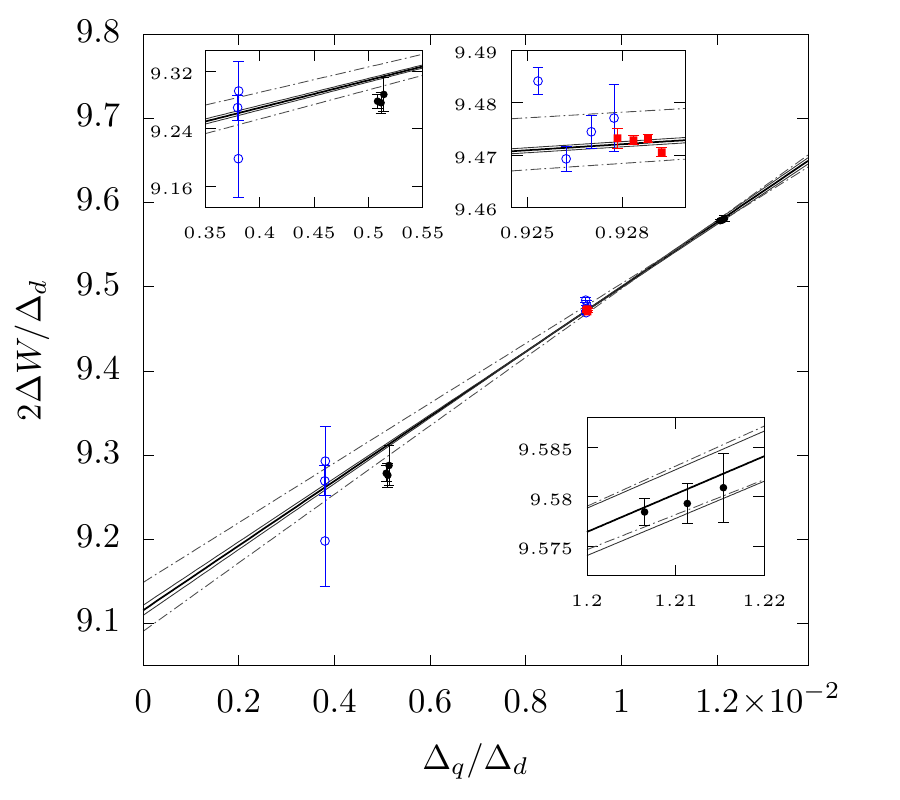}
    \caption{(color online) Core polarizability analysis including non-adiabatic corrections.
    As in Fig.~\ref{fig:adiabatic}, solid black circles are measurements from the present work, open blue points are from Lee~{\em et al.}~\protect\cite{Lee2016}, and solid red squares are from Moore~{\em et al.}
    \protect\cite{Moore20} 
    The axis quantities now include the non-adiabatic correction factors $k_d$
    and $k_q$, as discussed in the text. Quantities are in atomic units, $a_0^3$ (vertical) and $a_0^{-2}$ (horizontal).
    The heavy solid line is a fit to all the data. The thin solid lines
    illustrate the uncertainty in the fit, and the dashed lines show
    the confidence range using only the Berl and Lee data. The 
    uncertainties here do not include 
    the uncertainties of the non-adiabatic coefficients $k_i$.}
    \label{fig:nonadiabatic}
\end{figure}

Since we measure energy differences $\Delta W$, we again use Eq.~\eqref{DWadb}, but the definitions of $\Delta_d$ and $\Delta_q$ now include $k_{d, n\ell}$ and $k_{q, n\ell'}$ and are given by
\begin{equation} \label{deltad}
\Delta_d \equiv k_{d,n\ell} \avg{\frac{1}{r^{4}}}_{n\ell}
- k_{d,n\ell'} \avg{\frac{1}{r^{4}}}_{n\ell'}
\end{equation}
and
\begin{equation} \label{deltaq}
\Delta_q \equiv k_{q,n\ell} \avg{\frac{1}{r^{6}}}_{n\ell}
- k_{q,n\ell'} \avg{\frac{1}{r^{6}}}_{n\ell'}.
\end{equation}
As before, we plot $2\Delta W/\Delta_d$ vs.\
$\Delta_q/\Delta_d$, for $(\ell,\ell')$ pairs $(4,5)$ and $(5,6)$, shown as the solid circles in 
Fig.~\ref{fig:nonadiabatic}. 
Open circles are again data from Lee {\em et al.}~\cite{Lee2016},
and closed squares are from Moore {\em et al.}~\cite{Moore20}, with the Lee data plotted
as $2W/(k_d\langle r^{-4}\rangle)$ vs.\ 
$(k_q \langle r^{-6}\rangle)/(k_d \langle r^{-4}\rangle)$. The heavy line is
a fit to all the data, while the thin and dashed lines show the confidence
intervals with and without the Moore data, respectively. 
The fit results are listed in Table~\ref{tab:nonadiabatic},
with the first values in parentheses indicating the estimated uncertainties.

These fits do not account for the uncertainties in $k_d$ and $k_q$ themselves. To do so, 
we redo the analysis as the $W_I$ and $W_{Iq}$ parameters are varied independently
across their uncertainty ranges. The resulting changes in polarizabilities are indicated in
Table~\ref{tab:nonadiabatic} by the second value in parentheses. We take the total
uncertainty as the quadrature sum of the two values, leading to final results of
$\alpha_d = 9.12(3)$, $\alpha_q = 38.1(6)$ for the Berl and Lee data only, and
$\alpha_d = 9.116(9)$, $\alpha_q = 38.4(6)$ when the Moore data is included. The value of $\alpha_d$ is only 0.3\% different from its adiabatic value, almost in agreement with the adiabatic expansion model, which predicts no change \cite{hahn, snow}. However, $\alpha_q$ is almost double its adiabatic value, due primarily to $k_q$.

\begin{table}
	\caption{\label{tab:nonadiabatic}%
		Calculated values of the dipole and quadrupole polarizabilities, incorporating non-adiabatic
		corrections. Data sets are the same as in Table \protect\ref{tab:adiabatic}. Here the first
		value in parentheses is the esimtated error from the linear fit, and the second value is
		the estimated error from the non-adiabatic corrections. The bottom
		row shows our final estimated values with uncertainties, taken from the fit will all three data sets.
	}
	\begin{ruledtabular}
		\begin{tabular}{l|ccc}
			Data sets & $\alpha_d^{(a)}$ & $\alpha_q^{(a)}$ &  $\chi^2$/dof \\ \hline
			Berl & 9.060(19)(10) & 42.9(1.6)(6.3) & 0.11 \\
			Berl, Lee & 9.120(29)(7) & 38.1(2.6)(5.6) & 4.4 \\
			Berl, Lee, Moore & 9.116(6)(7) & 38.4(0.7)(5.6) & 3.7 \\ \hline
			Final values & 9.116(9) & 38.4(6) &
		\end{tabular}
	\end{ruledtabular}
\end{table}

Our results can be compared to previous theoretical estimates summarized in
Table~\ref{tab:theory}. We find good agreement with the most recent
results of \cite{Safronova2011}. This consistency resolves the large 
discrepancy between theory and the adiabatic $\alpha_q$ value reported in \cite{Lee2016}. 

\begin{table}[b]
	\centering
	\caption{Theoretical estimates of the core polarizability paramters.}
	\label{tab:theory}
	\begin{ruledtabular}
	\begin{tabular}{cll}
		~Ref.~ & $\alpha_d$~(au) & $\alpha_q$~(au) \\ \hline
		\protect\cite{Safronova2011} & 9.1 & 35.4 \\
		\protect\cite{johnson_electric-dipole_1983} & 9.076 & 35.41  \\
		\protect\cite{heinrichs_simple_1970} & 10.22 &   \\
		\protect\cite{sternheimer_quadrupole_1970} & & 38.43 
		\end{tabular}
	\end{ruledtabular}
\end{table}

Although we measure transition frequencies, we can use the extracted polarizabilities to calculate the absolute energy of the Rydberg states, and thus obtain the quantum defects. For this, we use Eq.~\eqref{Wadiab} and the adiabatic polarizability values $\alpha_d^{(a)}$ and $\alpha_q^{(a)}$, 
since that avoids the uncertainty in the non-adiabatic correction factors. 
The quantum defects are then found by setting
\begin{equation}
W_{n\ell} = \frac{R}{n^2}-\frac{R}{(n-\delta_{n\ell})^2}.
\end{equation}
We use the Ritz expansion \cite{Drake1990}
\begin{equation}
    \label{eq:ritz}
    \delta (n) = \delta_0 + \frac{\delta_2}{(n-\delta_0)^2}. 
\end{equation}
By expanding both Eq.~\eqref{Wadiab} and Eq.~\eqref{eq:ritz} in powers of $1/n$ and matching 
coefficients, we have
\begin{equation}
    \delta_{0,\ell} = \frac{1}{R} \left[ \frac{12(2\ell-2)!}{(2\ell+3)!} \alpha_d^{(a)}
                + \frac{560(2\ell-4)!}{(2\ell+5)!}\alpha_q^{(a)}\right]
\end{equation}  
and
\begin{align}
    \delta_{2,\ell} = & -2\delta_0^3 - \frac{1}{R} \left\{
            \frac{4\ell(\ell+1)(2\ell-2)!}{(2\ell+3)!} \alpha_d^{(a)} \right. \nonumber \\
           & + \left. \frac{480(2\ell-4)!}{(2\ell+5)!}
           \left[\ell(\ell+1) - \frac{5}{6}\right]\alpha_q^{(a)}\right\}.
\end{align}  
We use these expressions and the $\alpha_i^{(a)}$ values calculated with the Moore data to find the
results listed in Table~\ref{tab:defects}. The $g$-state values can be compared to 
$\delta_0 = 0.004 00(2)$, $\delta_2 = -0.018(15)$ obtained by Lee \cite{Lee2016} and
$\delta_0 = 0.003 999 (2)$, $\delta_2 = -0.020(2)$ obtained by Moore \cite{Moore20}.
The source of this discrepancy is likely related to the moderate inconsistencies
of the measurements noted in Table~\ref{tab:adiabatic}. We expect these inconsistencies
to be resolved with further investigations.

\begin{table}
	\centering
	\caption{Quantum defect Ritz expansion coefficients of Eq.~\protect\eqref{eq:ritz}.}
	\label{tab:defects}
	\begin{ruledtabular}
		\begin{tabular}{cll}
			$\ell$  & $\delta_0$ & $\delta_2$ \\ \hline
			$g$ & 0.004 007(5) & -0.027 42(6) \\
			$h$ & 0.001 423(1) & -0.014 38(2) \\
			$i$ & 0.000 607 4(4) & -0.008 550(8) 
		\end{tabular}
	\end{ruledtabular}
\end{table}

\section{Conclusions}

The measurements reported here provide a new set of constraints on the
core polarizabilites of Rb atoms, based on relatively low $n$ values. 
Together with new high-$n$ results from
Moore {\em et al.}\ \cite{Moore20}, the precisions of $\alpha_d$ and $\alpha_q$ are
improved by a factor of four compared to the previous work of Lee~{\em et al.}\ \cite{Lee2016}. In addition, we point out that non-adiabatic effects have
a significant impact on the value of the quadrupole polarizability $\alpha_q$, 
which brings the experimental results into line with theory.

We can consider methods to further improve the measurements. Since uncertainty
in the non-adiabatic corrections is significant, it would be helpful to determine them
with a more
sophisticated atomic structure calculation, compared to the empirical approach
described here. If such a calculation can be performed, then reducing the measurement
uncertainties would also be useful. A straightforward improvement would be to
add electrodes to the apparatus to allow control of the transverse electric field,
so that dc Stark shifts can be further reduced. 

Extending the measurements to even higher $\ell$ would provide a useful test of 
the core polarization model, and help identify any penetration shifts in the 
$g$-states. However, this is challenging because the signal-to-noise ratio on the $nf \rightarrow nj$ transition would be low in our existing apparatus. Further, the decreasing value of $\Delta W$ makes the relative frequency uncertainty more significant. A different approach would be to perform absolute spectroscopy of the $nf$ state so that the energy shifts relative to hydrogen of the $ng$, $nh$ and $ni$ states could be used independently. We cannot carry out such spectroscopy with our current apparatus: although precise spectroscopy of the $nd$ states is available \cite{Wenhui2003}, at low $n$ values the $nd-nf$ frequency intervals are too large to access with our microwave technology.

We expect that the improved
core polarizability values determined in this work will be useful for precision measurements such 
as atomic clocks and tune-out spectroscopy. 
In regards to our own interest in tune-out spectroscopy, the core polarizability was
a source of uncertainty in the determination of the ratio
of the $5p_{3/2}$ to $5p_{1/2}$ dipole matrix elements. The original analysis in
\cite{Leonard2015} used $\alpha_d = 9.08(10)$ au. Using the value $9.116(9)$ 
determined here, we 
find that the ratio is slightly reduced, from 1.99 217(3) to 1.992~15(3). 
We hope that further
improvements will allow us to reduce the uncertainty in this value and to
better constrain other dipole matrix elements of Rb as well \cite{Leonard2015}.

\begin{acknowledgments}
	This work was supported by National Science Foundation (grant Number PHY-1607571) and the Air Force Office of Scientific Research (grant Number FA9550-14-1-0288). J. Nunkaew was supported by the Thailand Center of Excellence in Physics (ThEP-61-PHY-MU3). It is a pleasure to acknowledge Adam Fallon and Safra Niyaz for useful discussions.
\end{acknowledgments}

\nocite{*}

\bibliography{main}

\end{document}